# Making Sense out of a Jungle of JavaScript Frameworks
## Towards a Practitioner-friendly Comparative Analysis

Daniel Graziotin, Pekka Abrahamsson

Free University of Bozen-Bolzano, Italy
`{daniel.graziotin, pekka.abrahamsson}@unibz.it`

**Abstract.** The field of Web development is entering the HTML5 and CSS3 era and JavaScript is becoming increasingly influential. A large number of JavaScript frameworks have been recently promoted. Practitioners applying the latest technologies need to choose a suitable JavaScript framework (JSF) in order to abstract the frustrating and complicated coding steps and to provide a cross-browser compatibility. Apart from benchmark suites and recommendation from experts, there is little research helping practitioners to select the most suitable JSF to a given situation. The few proposals employ software metrics on the JSF, but practitioners are driven by different concerns when choosing a JSF. As an answer to the critical needs, this paper is a call for action. It proposes a research design towards a comparative analysis framework of JSF, which merges researcher needs and practitioner needs.

**Keywords:** Web Development, JavaScript Framework, Comparative Analysis.

## 1   Introduction

Technologies like HTML5, CSS3, and JavaScript (JS) are maturing in a way that it is possible to substitute entire Desktop applications with counterparts running in a Web browser. Innovation is certainly not stopping here. Many argue that it is necessary for the industry to follow the momentum. JS is the most popular programming language for the browser [5]. Although it is possible to write pure JS code while constructing websites and Web applications, this is typically avoided with a JavaScript Framework (JSF). A JSF should abstract the longest and complex operations, ensure cross-browser compatibility, and speed up program comprehension and software development.

As of today, according to Jster.net, thousands of JS libraries are available for different purposes. Examples include *jQuery*, *Backbone.js*, *YUI*. When developing a Web application it is necessary to choose which framework to apply [2] as soon as possible as it introduces bindings and constraints. Software developers face difficulties when evaluating a JSF. Specialized websites like StackOverflow.com are full of beginner's questions on the topic[1].

---

[1] For example, http://fur.ly/9fp8

The focus of existing research is on the complexity and the quality of JS source-code (e.g. [2]). Software benchmarks lead on-line comparisons, to the point that Web browser vendors claim superior performance over the competitors by benchmarking performance on running JS code [4]. While benchmarks are able to measure different aspects of performance, their results may not be representative of real Web sites at all [4].

We note that practitioners seems interested in different aspects than those of academic research. For example, the Wikipedia page <http://en.wikipedia.org/wiki/Comparison_of_JavaScript_frameworks> compares 22 JSF without considering software metrics at all. Some of the criteria are the age of the latest release, the size of the JSF, the license, presence of features (e.g., Ajax, JSON Data Retrieval) and the browser support. Additionally, we note that the ability to obtain a JSF by expressing the concerns to be solved also seems useful for practitioners. Jster.net is an online catalog of JSF and libraries, where each project can be reached through semantic tags related to concerns (e.g., DOM traversing, Math Libraries, Routing, and UI Components).

It appears that the research interests in the academia are diverging from practitioners' interests. While this is not entirely uncommon, for this end, we call for action and propose a research design towards a comparative analysis framework of JSF. The resulting comparison framework will combine researcher's interests with the practitioner's interests in order to meet the best of the two worlds. We aim to expand an already proposed academic solution consisting of software metrics of JSF [2] with practitioner-related concerns.

## 2      Related Work

As far as we know, the literature consists in a single proposal. Gizas et al. [2] have attempted to compare six JavaScript Frameworks (JSF) using software metrics on the frameworks. They compare ExtJS, Dojo, jQuery, MooTools, Prototype, and YUI. They evaluate what they describe as the core version of each framework, namely DOM manipulation, selectors, Ajax functionalities, form elements, functions for event handling, and compatibility support. They test different aspects of quality, validation, and performance. The *quality* is expressed in terms of size metrics – i.e., statements, lines, comments, and related ratios, complexity metrics – i.e., branches, depth, McCabe's Cyclomatic Complexity, and maintainability metrics – i.e., Halstead metrics and Maintainability Index. The *validation* tests were performed by using the tools JavaScript Lint and Yasca. The *performance* tests were measurement of the execution time of the JSF with SlickSpeed Selectors test framework. The proposal is recent, in the form of a short paper.

## 3      Research Design

The proposal by Gizas, et al. [2] will be extended in two different measurement directions: one related to research and the other one to practitioners. (1) Their proposed metrics on code validation, quality, and performance will be employed on the JSF as they suggest, but also on the same Web application implemented with the different JSF. (2) Measurements related to practitioner concerns will emerge from in-field studies and interviews of developers. The GQM method [1] will be employed to find the most appropriate metrics to represent the practitioner concerns.

### 3.1 Pilot Study

We contacted four front-end Web developers to obtain their views on how to choose a JSF. The discussions were related to the criteria employed when choosing a new JSF or how a currently employed JSF was chosen. The preliminary results support the expected divergence between the proposed software metrics and the practitioner's criteria when selecting a JSF.

Three criteria were mentioned by all the participants: adequacy of the documentation, community participation, and "code less, do more" factor – i.e., the pragmatics of a JSF. Other emerged concerns are the maturity of the JSF and the frequency of the updates - i.e., its "freshness". How a JSF fulfills these concerns is subjectively perceived by inspecting the source code, code examples, and the documentation.

When asked about the metrics proposed by Gizas et al. [2], the respondents showed mild interest measurements of performance and admitted having a poor understanding of the other metrics. We were recommended to perform measurements on the same software project implemented using different JSF instead of measuring the JSF alone. A suitable project for this end is TodoMVC [3]. It enables practitioners to study and compare MV* (Model-View-Anything) JSF through source-code inspection of the same TODO-list Web application, developed by experienced Web developers employing their favorite JSF. TodoMVC provides a rigorous set of requirements, HTML/CSS templates, coding style and other specifications[2]. In order to be accepted in the TodoMVC catalog, the applications are first reviewed by the project leaders and then by the open-source community.

The participants suggested some measurements that would be beneficial for them. For example, the ratio of answers over questions related to a JSF on StackOverflow is perceived as being representative of the community involvement while traditional measurements like the frequency of commits in the version control system of the JSF represents both the community participation and the freshness of the JSF.

### 3.2 Proposed Framework

A high level view of the comparison framework is represented in Figure 1. The current proposal was born after the analysis of the pilot study data. The framework is organized in two layers, one related to research and the other related to practitioners. The blue boxes are the categories of metrics proposed by Gizas et al. [2] while the orange boxes are the extensions suggested by this study.

Each JSF will be measured using measurements relevant for academia and practitioners. Empirical data coming from the corresponding TodoMVC project will enforce the theoretical claims (thus, the blue boxes are surrounded by orange borders). In-field studies will improve the practitioner area of the framework. The GQM model will be employed to find the most appropriate metrics to represent the practitioner needs.

---

[2] Complete TodoMVC instructions and specifications: http://fur.ly/9fp7

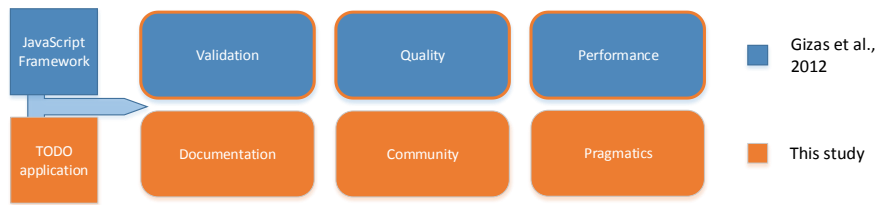

**Fig. 1.** Comparison Framework

## 4      Conclusion

We are entering a new era of Web Development, in which JavaScript (JS) is becoming more and more crucial for the Information Technology industry. So far, research interests have not been aimed at supporting the task of finding a suitable JavaScript framework (JSF) to improve the current state of Web development.

We set a call for action in this paper. We presented a research design towards a comparative analysis framework of JSF suitable for researchers and practitioners. The framework will extend a recent proposal to analyze JSF technically, using software metrics. These metrics will also be collected on the same software product produced using the different JSF. Empirical data from practitioners will be collected to understand and validate what are their needs when choosing a JSF. Therefore, research-related metrics will be complemented by practitioners-friendly metrics in a modern, updated database of JSF.

The resulting comparison framework will be a step forward in conciliating software engineering research and practitioners of software development. It will allow a quick selection of a JSF, thus saving time and resources of Web development firms.